\documentclass[
 reprint,
 amsmath,amssymb,
 aps,
]{revtex4-2}

\usepackage{graphicx}
\usepackage{dcolumn}
\usepackage{bm}
\usepackage{hyperref}
\usepackage[utf8x]{inputenc}

\begin{document}

\title{Statistical mechanics of the minimum vertex cover problem in stochastic block models}

\author{Masato Suzuki}
\author{Yoshiyuki Kabashima}
\affiliation{Department of Mathematical and Computing Science, Tokyo Institute
of Technology, 2-12-1 Ookayama, Meguro-ku, Tokyo 152-8552, Japan}

\date{\today}

\begin{abstract}
The minimum vertex cover (Min-VC) problem is a well-known NP-hard problem. Earlier studies illustrate that the problem defined over the Erd\"{o}s-R\'{e}nyi random graph with a mean degree $c$ exhibits computational difficulty in searching the Min-VC set above a critical point $c = e = 2.718 \ldots$. Here, we address how this difficulty is influenced by the mesoscopic structures of graphs. For this, we evaluate the critical condition of difficulty for the stochastic block model. We perform a detailed examination of the specific cases of two equal-size communities characterized by in- and out- degrees, which are denoted by $c_{\rm in}$ and $c_{\rm out}$, respectively. Our analysis based on the cavity method indicates that the solution search becomes difficult when $c_{\rm in }+c_{\rm out} > e$, but becomes easy again when $c_{\text{out}}$ is sufficiently larger than $c_{\mathrm{in}}$ in the region $c_{\rm out}>e$. Experiments based on various search algorithms support the theoretical prediction. 
\end{abstract}

\maketitle

\section{\label{sec:1}Introduction}
The minimum vertex cover problem (Min-VCP) is a well-known combinatorial optimization problem. Given a graph, the task of Min-VCP is to obtain a vertex set of minimum size, such that every edge in the graph is connected to at least one vertex in the set. Despite the simplicity of its expression, this problem is related to many real-world problems including monitoring internet traffic \cite{breitbart2001efficiently}, preventing denial-of-service attacks \cite{park2001effectiveness}, creating immunization strategies in networks \cite{gomez2006immunization}, solving the placement problem \cite{weigt2000number}, and dealing with data aggregation \cite{kavalci2014distributed}.In addition, Min-VCP is considered a representative computationally difficult problem that belongs to the class NP-hard. This means that, in the worst case, all known algorithms require an exponentially long time with respect to the number of vertices not only to find a solution but also to verify its validity.

Conventional theory in computer science determines the computational difficulty mainly in the worst case. However, characterizing the difficulty for a typical case is also essential for various practical situations. Recent studies in statistical mechanics carried out the typical case analysis of Min-VCP for the Erd\"{o}s-R\'{e}nyi (ER) random graph ensemble  \cite{jin2014statistical}, \cite{hartmann2006phase}, \cite{weigt2006message}. Based on the analysis using the replica and cavity methods, these studies reported that searching for the correct solution typically becomes computationally difficult when the mean degree $c$ exceeds $e = 2.718 \ldots$, while the solution can be obtained easily for $c < e$. The ER model (ERM) is simple and suitable as the initial step to analyze a typical case. Nonetheless, extending the result to more general ensembles is required since
the graphs in real world exhibit several mesoscopic structures, which are absent in ERM.

Based on the above perspective, this study addresses how the critical condition of computational difficulty is influenced when non-trivial structures are introduced into graphs. As a simple but non-trivial example, we examine the Min-VCP defined over the stochastic block model (SBM), which is a generalization of ERM often employed for the community detection problem \cite{abbe2017community}. Particularly focusing on the equal size two community cases, which can be characterized by two parameters, the mean in- and out-degrees, denoted by $c_{\text{in}}$ and $c_{\text{out}}$ respectively, we will show that the solution search becomes difficult when $c_{\text{in}} + c_{\text{out}} > e$, but becomes easy again when $c_{\rm out}$ is sufficiently larger than $c_{\rm in}$ in the region $c_{\rm out} > e$. This indicates that mesoscopic structures such as ``communities'' in graphs strongly influence the computational difficulty.

The remainder of this paper is organized as follows. In the next section, we formulate the problem. In section \ref{sec:3}, we analyze the Min-VCP defined over SBM using the cavity method, which yields the critical condition of the  computational difficulty as a function of $c_{\text{in}}$ and $c_{\text{out}}$ in typical cases. The theoretical prediction is then tested by numerical experiments in section \ref{sec:4}. The final section is devoted to the summary of our results.

\section{\label{sec:2}Definitions}

\subsection{\label{subsec:2-1}Minimum vertex cover problem}

We consider a graph $G$ with $N$ vertices $V = \{1, 2, ..., N\}$ and a set of edges $E$. A vertex cover (VC) $V_{\mathrm{VC}}$ of the graph $G$ is a subset of the vertices $V$ of the graph $G$ which includes at least one vertex of every edge in $E$. Then, Min-VCP is the problem of finding $V_{\rm VC}$ such that the number of elements in the VC is minimized.

Min-VCP is a well-known NP-hard problem. However, if the graph is bipartite, this problem is equivalent to the maximum matching problem, which belongs to the class P, according to K\"{o}nig's theorem \cite{bondy1976graph}. This indicates that the computational difficulty can vary depending on the graph structure of the problem. Our main purpose is to clarify how this change is characterized for the typical case.

\subsection{\label{subsec:2-2}Random graph models}

ERM is a random graph model constructed by connecting every pair of vertices $(i, j) \in V \times V$ independently with probability $p \in [0, 1]$. The degree distribution of ERM asymptotically converges to the Poisson distribution $P(d) = e^{-c}c^k / k!$, where $c := (N-1)p$ is the mean degree. Therefore, ERM is also called a Poisson-distributed graph.

On the other hand, SBM is a random graph model generalized from ERM to describe a community (or cluster) structure. For generating a sample of SBM, a hidden label $z_i \in \{1,\ldots, K\} \ (i=1,2,\ldots, N)$ is firstly assigned to each of the $N$ vertices with a categorical distribution $\pi = \{\pi_1, \pi_2\, ..., \pi_K\}$. Given a set of the community labels $\{z_i\}$, every pair of vertices $(i, j) \in V \times V$ is connected independently with probability $\Theta_{z_i, z_j}$, where $\Theta \in [0, 1]^{K \times K}$ is an affinity matrix with the property $\Theta_{z,z^\prime} = \Theta_{z^\prime, z}$. Then, the probability that a node of community $z$ has $d$ degrees to community $z'$ is given by
\begin{align}
\label{eq:degreedist}
P_{z, z'}(d) &:= e^{-c_{z, z'}} \frac{c_{z, z'}^d} {d!}, \\
\label{eq:meandegree}
c_{z, z'} &:= \begin{cases} (N \pi_{z} - 1) \Theta_{z, z}, & z = z' \\ N \pi_{z'} \Theta_{z, z'}, & z \neq z' \end{cases}
\end{align}
where $c_{z, z'}$ is the mean degree from community $z$ to community $z'$. We assume that $\Theta_{z, z'}$ is set such that $c_{z, z'}$ is $O(1)$.

By controlling $\pi$ and $\Theta$, one can implement various community structures in SBM. As a particular case, we consider the SBM model with $\pi_z = 1 / K$ and $\Theta_{z,z'} = \begin{cases} p_{\mathrm{in}} & z = z' \\ p_{\mathrm{out}} & z \neq z' \end{cases}$ with $p_{\mathrm{in}}, p_{\mathrm{out}} \in [0, 1]$, namely, every vertex is assigned to each community uniformly. In this case, the mean degree of intra-community connections is $c_{\mathrm{in}} := \left(\frac N K - 1 \right) p_{\mathrm{in}}$ and that of inter-community connections is $c_{\mathrm{out}} := \frac N K p_{\mathrm{out}}$. This is called the symmetric stochastic block model (SSBM) \cite{abbe2017community}.

\section{\label{sec:3}Analytical framework}

This section introduces the analytical framework used in our study.

\subsection{\label{subsec:3-1}Statistical mechanical formulation}
 
Our analysis is based on the description of Min-VCP in terms of statistical mechanics \cite{jin2014statistical}. First, for a vertex cover $V_{\mathrm{VC}} \subset V$ of a graph $G = (V, E)$, we assign binary variables $x = \{x_i\}_{i \in V}$ to each vertex as follows.
\begin{equation}
\label{eq:1}
x_i := \begin{cases}
1 & v_i \in V_{\mathrm{VC}} \\
0 & v_i \notin V_{\mathrm{VC}}.
\end{cases}
\end{equation}
This enables us to express the size of $V_{\rm VC}$ simply as 
\begin{equation}
\label{eq:2}
H(x) := \sum_{i \in V} x_i.
\end{equation}

On the other hand, the constraint that every edge must be connected to at least one element of $V_{\rm VC}$ is expressed by
\begin{equation}
\label{eq:constraint}
\Delta(x) := \prod_{(i,j)\in E} (1-(1-x_i)(1-x_j))
\end{equation}
which returns unity if the constraint is satisfied and vanishes, otherwise. Combining these leads to the Boltzmann distribution
\begin{eqnarray}
\label{eq:3}
P(x) &=& \frac 1 {Z(\beta)} e^{-\beta H(x)} \Delta(x), \\
\label{eq:4}
Z(\beta) &=& \sum_{x} e^{-\beta H(x)} \Delta(x)
\end{eqnarray}
which yields the uniform distribution of the Min-VC sets in the limit of $\beta \to \infty$.

Based on this formulation, the internal energy per variable can be evaluated as
\begin{equation}
\label{eq:5}
\nu(\beta) = \frac{1}{N} \sum_{x} P(x) H(x) = -\frac{\partial }{\partial \beta} \frac{1}{N} \ln Z(\beta).
\end{equation}
This is reduced to the ratio of the covered vertices of the Min-VC set as
\begin{equation}
\label{eq:6}
x_c =\frac{1}{N} \mathop{\rm min}_{V_{\rm vc}} \left \{ |V_{\rm vc} | \right \} = \lim_{\beta \to \infty} \nu(\beta) 
\end{equation}
in the limit of $\beta \to \infty$.

\subsection{\label{subsec:3-2}Cavity method}

We denote by $\partial i$ and $x_{\partial i}$ the set of nearest neighbors of a vertex $i$ and that of variables indexed by the neighbors, respectively. In addition, we introduce a distribution $p_{\partial i \to i} (x_{\partial i})$ as
\begin{equation}
\label{eq:cav1}
p_{\partial i \to i} (x_{\partial i}) = \frac 1 {Z_{\setminus i}} \sum_{x \setminus x_{\partial i}} e^{-\beta \sum_{k \neq i} x_k} \prod_{k, l \neq i} (1 - (1 - x_k) (1 - x_l))
\end{equation}
where $A \setminus B$ generally indicates exclusion of subset $B$ from set $A$, and $Z_{\setminus i}$ is the normalization constant. This distribution stands for the joint distribution of $x_{\partial i}$ for the $i$-cavity system that is defined by excluding vertex $i$ from the original system. We call $p_{\partial i \to i}(x_{\partial i})$ the joint cavity distribution. For any vertex $i$ of general graphs, the following identity holds between the marginal distribution of variable $x_i$ and the joint cavity distribution $p_{\partial i \to i} (x_{\partial i})$.
\begin{equation}
\label{eq:cav2}
\begin{split}
p_i(x_i)
&= \sum_{x \setminus \{x_i\}} P(x) \\
&\propto \sum_{x \setminus \{x_i\}} e^{-\beta x_i} \prod_{j \in \partial i} (1 - (1 - x_i) (1 - x_j)) \\
&\quad \times e^{-\beta \sum_{k \neq i} x_k} \prod_{k, l \neq i} (1 - (1 - x_k) (1 - x_l)) \\
&\propto \sum_{x \setminus \{x_i\}} e^{-\beta x_i} \prod_{j \in \partial i} (1 - (1 - x_i) (1 - x_j)) p_{\partial i \to i} (x_{\partial i}).
\end{split}
\end{equation}

Evaluating $p_{\partial i \to i} (x_{\partial i})$ is non-trivial for general graphs. However, when $G$ does not contain any cycles, it is decomposed as
\begin{equation}
\label{eq:cav3}
p_{\partial i \to i} (x_{\partial i}) = \prod_{j \in \partial i} p_{j \to i} (x_j)
\end{equation}
where $p_{j \to i}(x_j)$ represents the marginal distribution of $x_j$ in the $i$-cavity system, since the exclusion of vertex $i$ makes the components of $x_{\partial i}$ statistically independent of one another. This offers simplified expressions for the distributions as
\begin{eqnarray}
\label{eq:8}
p_i(x_i = 1) &=& \frac{e^{-\beta}}{e^{-\beta} + \prod_{j \in \partial i} p_{j \to i}(x_j = 1)}, \\
\label{eq:9}
p_i(x_i = 0) &=& 1 - p_i(x_i = 1).
\end{eqnarray}

Further, by considering the process of adding vertex $j$ to the $j$-cavity system and excluding vertex $i \in \partial j$, we can efficiently compute the marginal cavity distribution $p_{j \to i}(x_j)$ using a message passing algorithm as
\begin{eqnarray}
\label{eq:10}
p_{j \to i}(x_j = 1) &=& \frac{e^{-\beta}}{e^{-\beta} + \prod_{k \in \partial j \setminus i} p_{k \to j}(x_k = 1)}, \\
\label{eq:11}
p_{j \to i}(x_j = 0) &=& 1 - p_{j \to i}(x_j = 1).
\end{eqnarray}
This procedure for evaluating the marginal distributions $p_i(x_i)$ by efficiently computing the cavity distributions $p_{j \to i}(x_j)$ for cycle-free graphs is often termed \textit{belief propagation (BP)} \cite{pearl1988probabilistic}.

Two issues should be noted here. The first one is about the treatment of graphs containing cycles. When graphs contain cycles, Eq.(\ref{eq:cav3}) does not hold, and therefore, BP does not yield an exact assessment of the marginal distributions. However, it can still be employed as an approximate algorithm since Eq.(\ref{eq:10}) is performable even if there are cycles. In particular, such treatment is expected to offer a good approximation of an accurate solution for large sparse random graphs such as ERM and SBM since the typical length of cycles for these graphs diverges as $O(\ln N)$, and therefore, their influence becomes negligible as $N$ tends to infinity. This motivates us to employ BP for searching the Min-VC set, which corresponds to the Bethe-Peierls approximation known in physics.

The second point to note is that, in the limit of $\beta \to \infty$, the marginal and cavity distributions can be expressed in simpler forms as
\begin{eqnarray}
\label{eq:12}
p_i(x_i = 1) &=&
\begin{cases}
0 & \sum_{j \in \partial i} u_{j \to i} = 0 \\
\frac 1 2 & \sum_{j \in \partial i} u_{j \to i} = 1 \\
1 & \sum_{j \in \partial i} u_{j \to i} > 1
\end{cases},\\
\label{eq:13}
p_{j \to i}(x_j = 1) &=&
\begin{cases}
0 & \sum_{k \in \partial j \setminus i} u_{k \to j} = 0 \\
\frac 1 2 & \sum_{k \in \partial j \setminus i} u_{k \to j} = 1 \\
1 & \sum_{k \in \partial j \setminus i} u_{k \to j} > 1
\end{cases}
\end{eqnarray}
using binary messages $\{u_{j \to i}\}$ that are determined by a reduced expression of BP
\begin{equation}
\label{eq:14}
u_{j \to i} = \begin{cases} 1 & \sum_{k \in \partial j \setminus i} u_{k \to j} = 0\\ 0 & \sum_{k \in \partial j \setminus i} u_{k \to j} \geq 1 \end{cases}.
\end{equation}
This procedure is referred to as \textit{warning propagation (WP)}. A comprehensive graphical explanation for WP is given in \cite{hartmann2006phase}. After determining the messages by WP, the cover ratio is evaluated as
\begin{equation}
\label{eq:15}
x_{\mathrm{c}} = \frac 1 N \sum_{i = 1}^N p_i(x_i = 1).
\end{equation}

\subsection{\label{subsec:3-2-5}Density evolution}

So far, we have investigated how to obtain the Min-VC set for a given single sample of graphs. However, ignoring the influence of cycles in graphs, which is expected to be valid for $N \to \infty$ in ERM and SBM, makes it possible to characterize the typical properties of graph ensembles using the densities of messages. In addition, in the limit of $\beta \to \infty$, the densities are expressed in particularly simple forms as
\begin{equation}
\label{eq:de1}
R_{z, z'}(u) := (1 - \rho_{z, z'}) \delta (u) + \rho_{z, z'} \delta (u - 1)
\end{equation}
where $z$ and $z'$ denote the labels of communities and $\rho_{z, z'} \in [0, 1]$ is the probability that the binary message $u_{j \to i}$ sent from node $j$ of community $z$ to $i$ of $z'$ takes the value of unity.

For ERM, which corresponds to the single community case $(K = 1)$, Ref. \cite{weigt2006message} showed that $\rho_z = W(c)/c$, where $W(\cdot)$ is the Lambert-W function. On the other hand, in the case of SBM, handling WP as the elementary process provides a set of self-consistent equations to determine $P_z(u)$, which is often termed the density evolution (DE), as
\begin{align}
\label{eq:de2}
\rho_{z, z'}
&= \sum_{d_1, ..., d_K} \left(\prod_{z'' \neq z'} P_{z, z''}(d_{z''}) \right) Q_{z, z'} (d_{z'}) \nonumber \\
&\qquad \times \prod_{z''=1}^K (1 - \rho_{z'', z})^{d} \nonumber \\
&= \exp\left (- \sum_{z''=1}^K c_{z,z''} \rho_{z'', z} \right )
\end{align}
where $Q_{z, z'}(d) = (d+1) P_{z, z'}(d+1) / \sum_{d'=0}^\infty d' P_{z, z'}(d')$ is the probability that when an edge is randomly chosen, a terminal node $i$ of $z$ has remaining degree $d$. Note that Eq. (\ref{eq:de2}) is independent of $z'$. Thus for the simplicity of notation, we hereafter denote $\rho_{z,z^\prime}$ as $\rho_z$.

Similarly, we express the densities of marginal probabilities as
\begin{equation}
\label{eq:de4}
R_z(p) = \nu_z^{(0)} \delta (p) + \nu_z^{\left( \frac 1 2 \right)} \delta \left( p - \frac 1 2 \right) + \nu_z^{(1)} \delta (p - 1)
\end{equation}
where $\nu_z^{(p)} \in [0, 1]$ for $p \in \left\{0, \frac 1 2, 1 \right\}$ is the probability that $p_i(x_i=1)=p$ holds for a randomly chosen node $i$ of community $z$.

After determining $\rho_z$ from Eq. (\ref{eq:de2}), $\nu_z^{(p)}$ in Eq. (\ref{eq:de4}) is evaluated as
\begin{align}
\label{eq:de4-1}
\nu_z^{(0)} &= \sum_{d_1, ..., d_K} \left(\prod_{z'=1}^K P_{z, z'}(d)\right) \prod_{z'=1}^K (1 - \rho_{z'})^d \nonumber \\
&= \rho_z, \\
\label{eq:de4-2}
\nu_z^{\left(\frac 1 2\right)}
&= \sum_{d_1, ..., d_K} \left(\prod_{z'=1}^K P_{z, z'}(d)\right) \sum_{z'=1}^K d_{z'} (1 - \rho_{z'})^{d_{z'}-1} \rho_{z'} \nonumber \\
&\hspace{120pt} \times \prod_{z'' \neq z} (1 - \rho_{z''})^{d_{z''}} \nonumber \\
&= \sum_{z'=1}^K c_{z, z'} \rho_z \rho_{z'}, \\
\label{eq:de4-3}
\nu_z^{(1)} &= 1 - \nu_z^{(0)} - \nu_z^{\left(\frac 1 2\right)}.
\end{align}
This yields the typical cover ratio for SBM as
\begin{equation}
\label{eq:de3}
x_{\rm c} =\mathbb{E} \left[ \frac{1}{N} 
\mathop{\rm min}_{V_{VC}} \left \{
|V_{VC}| \right \} \right]
= \sum_{z=1}^K \pi_z \left( \nu_z^{(1)} + \frac 1 2 \nu_z^{\left(\frac 1 2\right)} \right)
\end{equation}
where $\mathbb{E}[\cdot]$ stands for the average with respect to the distribution of graphs of SBM. Eqs. (\ref{eq:de2}) and (\ref{eq:de3}) constitute the first contribution of this study.

\subsection{\label{subsec:3-3}Stability analysis}

In this subsection, we examine the critical points at which the stability of the fixed point solution of WP is broken using two analytical methods. In the following, we assume that the update scheme for WP is a random sequential update, where one directed edge $j\to i$ is picked up uniformly in a given graph at every time step and the binary message on it is updated according to Eq. (\ref{eq:14}).

\subsubsection{Linear stability analysis for DE}

We denote $\rho_z$ at the $t$-th step in DE by $\rho_z^{(t)}$. We now consider the process of picking up directed edge $j\to i$ at random from SBM at step $t$ of WP. In this process, the probability that the binary message $u_{j→i}$ sent from node $j$ of community $z$ to $i$ of $z'$ takes the value of zero at the next step $t+1$ is evaluated as
\begin{align}
\label{eq:ls1}
q_{z, z'} &= \sum_{d_1, ..., d_K} \left(\prod_{z'' \neq z'} P_{z, z''}(d_{z''}) \right)Q_{z, z'}(d_{z'}) \nonumber \\
&\quad \times \prod_{z'' = 1}^K (1 - (1 - \rho_{z''}^{(t)})^{d_{z''}}) \nonumber \\
&= \prod_{z'' = 1}^K (1 - \exp(-c_{z, {z''}} \rho_{z''}^{(t)})).
\end{align}

Similarly, the probability that the binary message $u_{j→i}$ sent from node $j$ of community $z$ to $i$ of $z'$ takes the value of unity at the next step $t+1$ is evaluated as
\begin{align}
\label{eq:ls2}
r_{z, z'} &= \sum_{d_1, ..., d_K} \left(\prod_{z'' \neq z'} P_{z, {z''}}(d_{z''}) \right)Q_{z, z'}(d_{z'}) \nonumber \\
&\quad \times \prod_{z'' = 1}^K (1 - \rho_{z''}^{(t)})^{d_{z''}} \nonumber \\
&= \prod_{z'' = 1}^K \exp(-c_{z, {z''}} \rho_{z''}^{(t)}).
\end{align}

Eqs. (\ref{eq:ls1}) and (\ref{eq:ls2}) are independent of $z'$. Hereafter, $q_{z,z'}$ and $r_{z, z'}$ are denoted as $q_z$ and $r_{z}$, respectively.

In SBM, the two probabilities $q_{z}$ and $r_{z}$ become independent of the value of message $u_{j\to i}$ at step $t$ as the size of graphs tends to infinity. This indicates that the difference between $\rho_z^{(t+1)}$ and $\rho_z^{(t)}$ is expressed as 
\begin{align}
\label{eq:19}
\rho_z^{(t+1)} - \rho_z^{(t)}
&\simeq \frac 1 {M_z} \left( - \frac{M_z \rho_z^{(t)}}{M} q_z + \frac{M_z(1 - \rho_z^{(t)})}{M} r_z \right) \nonumber \\
&= \frac 1 M \left( - \rho_z^{(t)} q_z + (1 - \rho_z^{(t)}) r_z \right)
\end{align}
where $M_z := N \pi_z \sum_{z' = 1}^K c_{z, z'}$ is the number of messages that are sent from the nodes of community $z$ and $M := \sum_{z=1}^K M_z$ is the number of all messages. As $M \to \infty$, Eq. (\ref{eq:19}) can be treated as a set of ordinary differential equations with the rescaled time $\tau = t / M (d \tau = d t / M = 1 / M)$ as
\begin{equation}
\label{eq:ls3}
\frac{d \rho_z} {d \tau}
= - \rho_z q_z + (1 - \rho_z) r_z
\end{equation}
This makes it possible to examine the linear stability of DE of WP for SBM numerically. This is the second contribution of this study.

Specifically for the case of SSBM with two communities, Eq.(\ref{eq:ls3}) can be expanded as
\begin{equation}
\label{eq:ssbm_rho_ev}
\begin{split}
\frac{d \rho} {d \tau}
&\simeq \binom
{ - \rho_1 + \exp(- c_{\text{in}} \rho_1 - c_{\text{out}} \rho_2)}
{ - \rho_2 + \exp(- c_{\text{in}} \rho_2 - c_{\text{out}} \rho_1)}
\end{split}
\end{equation}
where $\rho = (\rho_1, \rho_2)^{\top}$ and $\top$ denotes the operation of matrix/vector transpose. The Jacobian $J$ is then given as
\begin{align}
\label{eq:25}
J &= - I - A, \\
\label{eq:ls4}
A &= \left(
\begin{array}{r}
c_{\mathrm{in}} \exp(-c_{\mathrm{in}} \rho_1 - c_{\mathrm{out}} \rho_2) \\
c_{\mathrm{out}} \exp(-c_{\mathrm{in}} \rho_2 - c_{\mathrm{out}} \rho_1)
\end{array}
\right .
\nonumber
\\
& \left . \hspace{50pt}
\begin{array}{r}
c_{\mathrm{out}} \exp(-c_{\mathrm{in}} \rho_1 - c_{\mathrm{out}} \rho_2) \\
c_{\mathrm{in}} \exp(-c_{\mathrm{in}} \rho_2 - c_{\mathrm{out}} \rho_1)
\end{array}
\right ).
\end{align}
where $I$ is the identity matrix. Therefore, the linear stability around a fixed point solution $\hat \rho$ of Eq. (\ref{eq:de2}) can be assessed by examining $J$ at $\hat \rho$. If all eigenvalues of $J$ at $\hat \rho$ are smaller than zero, the solution is linearly stable with respect to DE.

\subsubsection{Bug proliferation analysis for WP}

DE macroscopically characterizes the solution of WP. However, even if DE converges to a stationary density, it does not necessarily imply the convergence of messages to a fixed point since the messages can continue to move while keeping the density stationary. In ERM, earlier studies \cite{jin2014statistical}, \cite{hartmann2006phase}, \cite{weigt2006message} reported that the cover ratio evaluated by WP coincides with that by other methods for $c < e$, but does not for $c > e$. This transition is accompanied by the instability of the fixed point of WP, which is regarded as a consequence of the replica symmetry breaking \cite{jin2014statistical}, \cite{hartmann2006phase}, \cite{weigt2006message}. To examine the possibility of such a transition for SBM, we generalize a method termed the \textit{bug proliferation analysis} that was introduced for ERM in \cite{weigt2006message}.

The analysis starts with the assumption that messages are at a fixed point of WP. At the fixed point, we flip one message $u_{j \to i}$ to the other value, called a bug, and examine whether the bugs proliferate or die out during iterations of WP.

\begin{figure}[t]
 \begin{minipage}{0.45\hsize}
  \begin{center}
   \includegraphics[width=0.9\hsize]{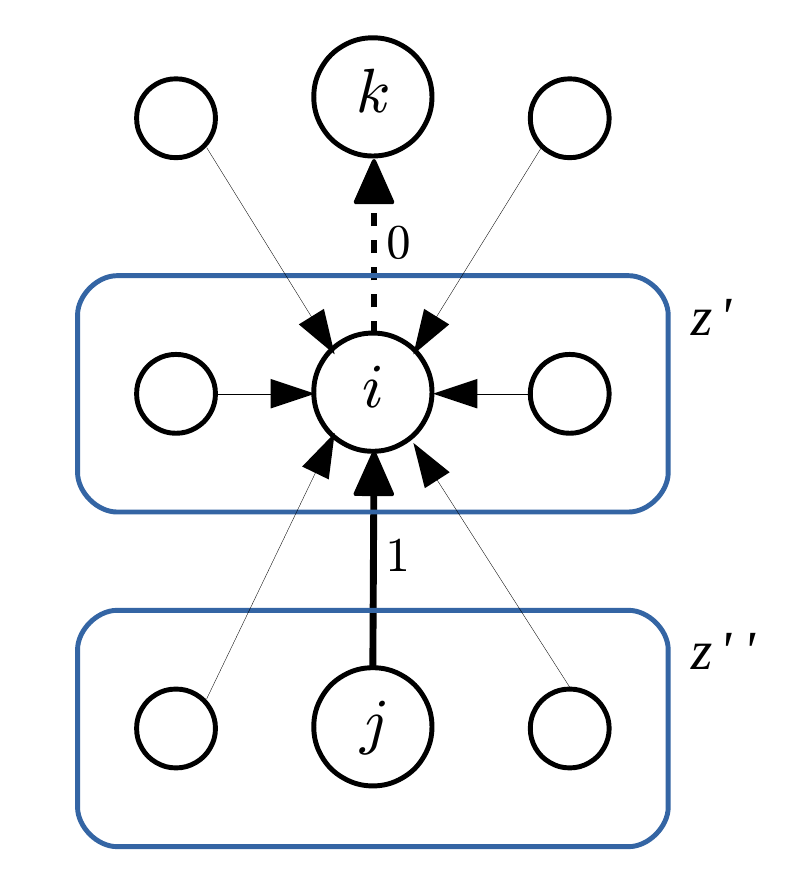}
   (a)
  \end{center}
 \end{minipage}
  \begin{minipage}{0.45\hsize}
  \begin{center}
   \includegraphics[width=0.9\hsize]{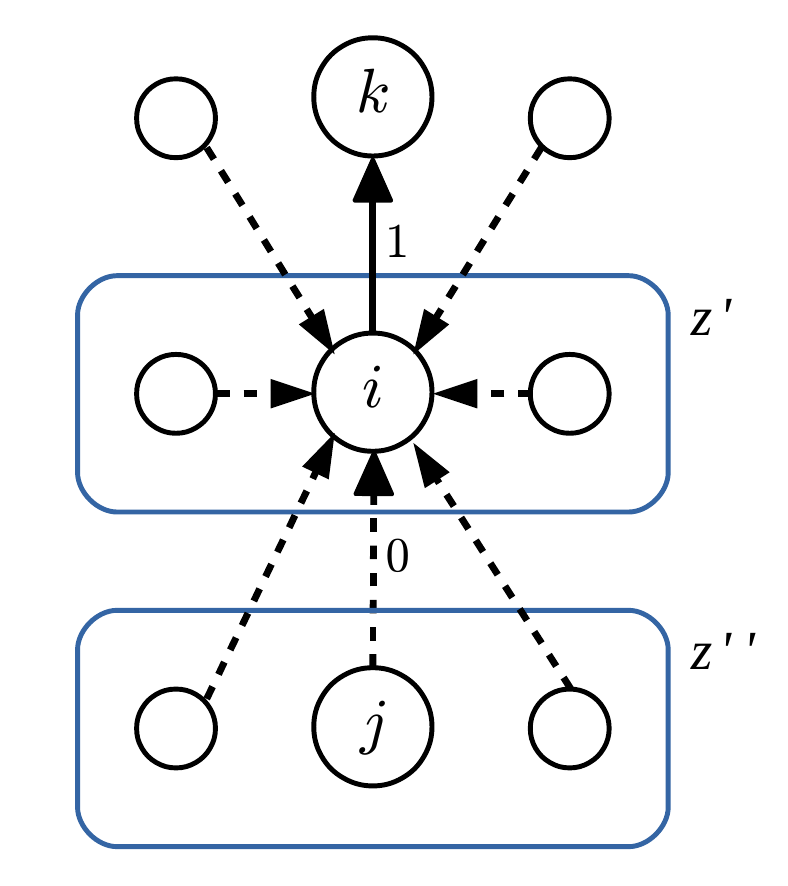}
   (b)
  \end{center}
 \end{minipage}
 \caption{When $u_{j\to i}=1$ is sent from node $j$ of community $z^{\prime \prime}$ to $i$ of $z^\prime$, every outgoing message $u_{i \to k}$ from $i$ to $k \in \partial i \setminus \{j\}$ is set to zero. Therefore, if an incoming message $u_{j \to i}$ is flipped from one to zero ((a) $\rightarrow$ (b)), each outgoing message $u_{i \to k}$ becomes a new bug if its value is changed to one by Eq. (\ref{eq:14}), that is, all incoming messages $u_{h \to i}$ from $h \in \partial i \setminus \{j, k\}$ to $i$ are zeros. On the other hand, if an incoming message $u_{j \to i}$ is flipped from zero to one ((b) $\rightarrow$ (a)), $u_{i \to k}$ becomes a new bug if its original value is one. This condition is equivalent to the previous one.}
 \label{fig:bug}
\end{figure}

Let us suppose that an incoming message $u_{j \to i}$ sent from node $j$ of community $z^{\prime\prime}$ to $i$ of $z^\prime$ is flipped. Then, the outgoing messages $u_{i \to k}$ from $i$ to $k \in \partial i\backslash \{j\}$ change to the other values if and only if all incoming messages from $h \in \partial i \backslash \{j,k \}$ to $i$ are zeros (Fig. \ref{fig:bug}).

Since node $i$ is selected randomly from $z'$, the expected number of the messages incoming to $z$ that are flipped by Eq. (\ref{eq:14}) due to a bug incoming from $z''$ to $z'$ is evaluated as
\begin{align}
\label{eq:bp1}
s_{z, z', z''} &= \sum_{d_1, ..., d_K} \left( \prod_{z^* \neq z'} P_{z, z^*} (d_{z^*}) \right) Q_{z, z'} (d_{z'}) \nonumber \\
&\quad \times d_{z'} \left( \prod_{z^* \neq z'} (1 - \rho_{z^*})^{d_{z^*}} \right) (1 - \rho_{z'})^{d_{z'} - 1}\nonumber \\
&= c_{z, z'} \prod_{z^* = 1}^K \exp(-c_{z, z^*} \rho_{z^*}).
\end{align}
Since Eq. (\ref{eq:bp1}) is independent of $z''$, we hereafter denote $s_{z, z', z''}$ as $s_{z, z'}$.

Next, we denote as $p_z^{(t)}(b_z)$ the probability that there are $b_z$ bugs incoming to community $z$ at the $t$-th step of the random sequential updates of WP. This probability is updated as
\begin{align}
\label{eq:bp4}
p_z^{(t+1)}(b_z)
&= p_z^{(t)}(b_z) - \frac{b_z}{M} p_z^{(t)}(b_z) + \frac{b_z + 1}{M} p_z^{(t)}(b_z + 1) \nonumber \\
&\qquad - \frac{s_{z, z} b_z}{M} p_z^{(t)}(b_z) + \frac{s_{z, z} (b_z - 1)}{M} p_z^{(t)}(b_z - 1) \nonumber \\
&\qquad - \sum_{z' \neq z} \sum_{b_{z'} = 0}^\infty \frac{s_{z, z'} b_{z'}}{M} p_{z'}^{(t)} (b_{z'}) p_z^{(t)} (b_z) \nonumber \\
&\qquad + \sum_{z' \neq z} \sum_{b_{z'} = 0}^\infty \frac{s_{z, z'} b_{z'}}{M} p_{z'}^{(t)} (b_{z'}) p_z^{(t)} (b_z - 1).
\end{align}
The second and third terms mean that a bug in the community $z$ is selected with the probability $b_z / M$ for each step, and is fixed. The fourth to seventh terms correspond to the case when a child of bugs which creates a new bug in the community $z$ is selected.

Similar to Eq. (\ref{eq:ls3}), in the limit $M \to \infty$, this can be rewritten as ordinal differential equation with the rescaled time $\tau = t/M$ as
\begin{align}
\label{eq:bp5}
\frac{d}{d\tau}p_z(b_z)
&= - b_z p_z(b_z) + (b_z - 1) p_z(b_z - 1) \nonumber \\
&\qquad - s_{z, z} b_z p_z(b_z) + s_{z, z} (b_z - 1) p_z(b_z - 1) \nonumber \\
&\qquad - \sum_{z' \neq z} \sum_{b_{z'} = 0}^\infty s_{z, z'} b_{z'} p_{z'} (b_{z'}) p_z (b_z) \nonumber \\
&\qquad + \sum_{z' \neq z} \sum_{b_{z'} = 0}^\infty s_{z, z'} b_{z'} p_{z'} (b_{z'}) p_z (b_z - 1).
\end{align}

Let us denote the average of $b_z$ at time $\tau$ as $\overline{b_z} := \sum_{b_z=0}^\infty b_z p_z(b_z)$. Eq. (\ref{eq:bp5}) indicates that $\overline{b_z}$ obeys ordinary differential equation 
\begin{align}
\label{eq:bp3}
\frac{d}{d\tau} \overline{b_z}
&= - \overline{b_z^2} + \overline{b_z (b_z - 1)} \nonumber \\
&\qquad - s_{z, z} \overline{b_z^2} + s_{z, z} \overline{b_z (b_z + 1)} \nonumber \\
&\qquad - \sum_{z' \neq z} s_{z, z'} \overline{b_{z'}} \overline{b_{z}} + \sum_{z' \neq z} s_{z, z'} \overline{b_{z'}} \overline{b_{z} + 1} \nonumber \\
&= - \overline{b_z} + \sum_{z' = 1}^K s_{z, z'} \overline{b_{z'}}.
\end{align}
which is the generalization of the bug proliferation analysis proposed for ERM in \cite{weigt2006message} to SBM. This is the third contribution of this study. 

In the case of SSBM with two communities, Eq. (\ref{eq:bp3}) is expressed in a simple form as
\begin{equation}
\label{eq:26}
\frac{d} {d \tau} \overline b = (-I + A) \cdot {\overline b}
\end{equation}
where $\overline b = (\overline b_1, \overline b_2)^{\top}$ and $A$ is introduced at Eq. (\ref{eq:ls4}). The corresponding Jacobian is then given by $-I + A$. Unless every eigenvalue of $- I + A$ at $\hat \rho$ has an absolute values smaller than zero, the fixed point of WP is unstable.

\section{\label{sec:4}Numerical results}

In this section, we compare the analysis in Sec. \ref{sec:3} with the results of numerical experiments for SSBM with two communities.

\subsection{Stability analysis for SSBM with two communities}

\begin{figure}[t]
    \centering
    \includegraphics[width=\hsize]{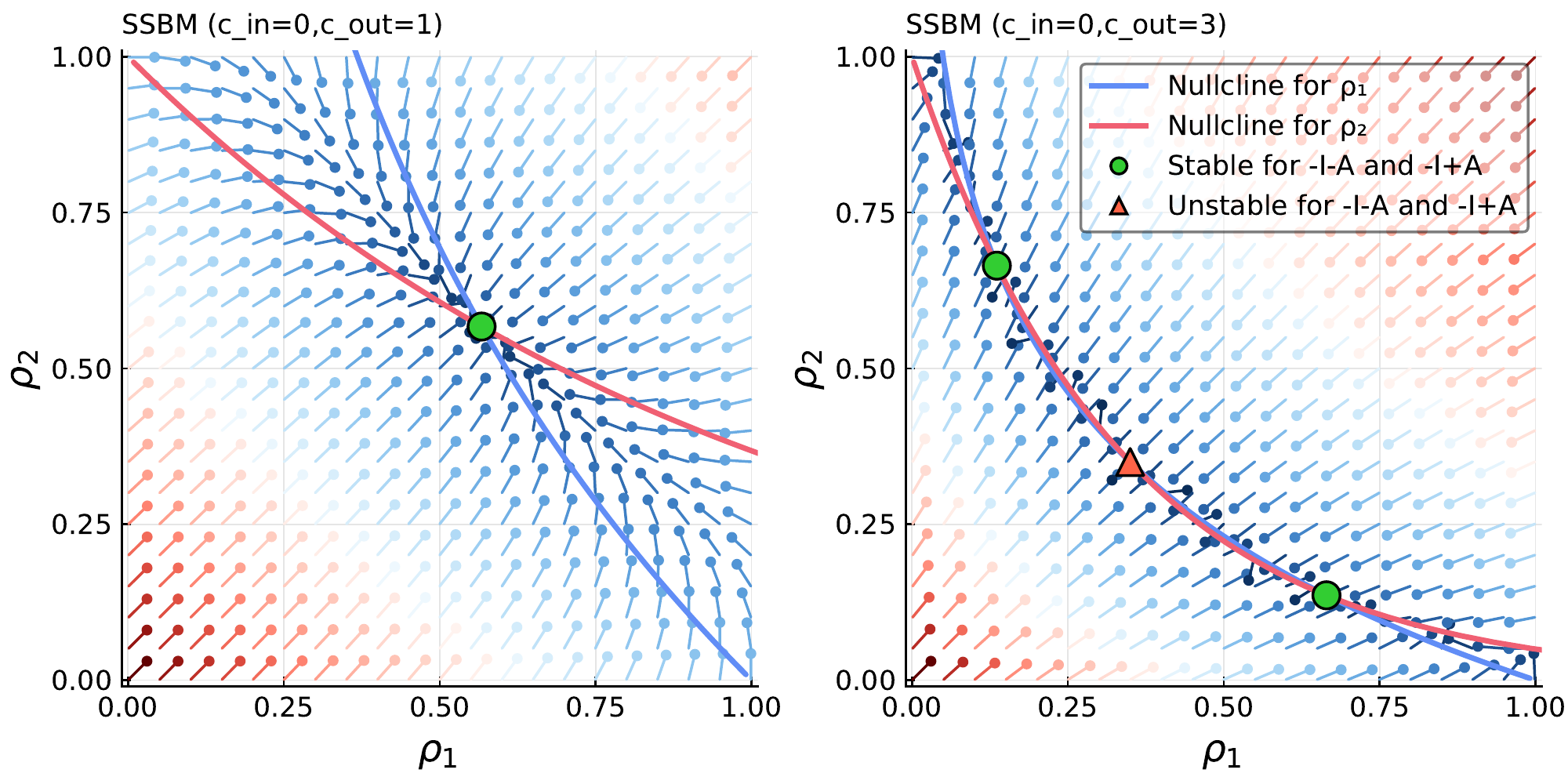}
    \caption{The phase planes of Eq.(\ref{eq:ssbm_rho_ev}) for SSBM with two communities. The mean degree of intra-community is fixed as $c_{\text{in}} = 0$. Solid curves are nullclines of the system, and each marker represents its fixed points. The color of these points indicate the stability evaluated by $- I - A$ (Eq.(\ref{eq:25})), that is, the linear stability of $\rho = (\rho_1, \rho_2)^{\top}$. The difference of the markers indicates the stability assessed by $- I + A$ (Eq.(\ref{eq:25})), that is, the microscopic stability with respect to WP.}
    \label{fig:phaseplain1}
\end{figure}

\begin{figure}[t]
    \centering
    \includegraphics[width=\hsize]{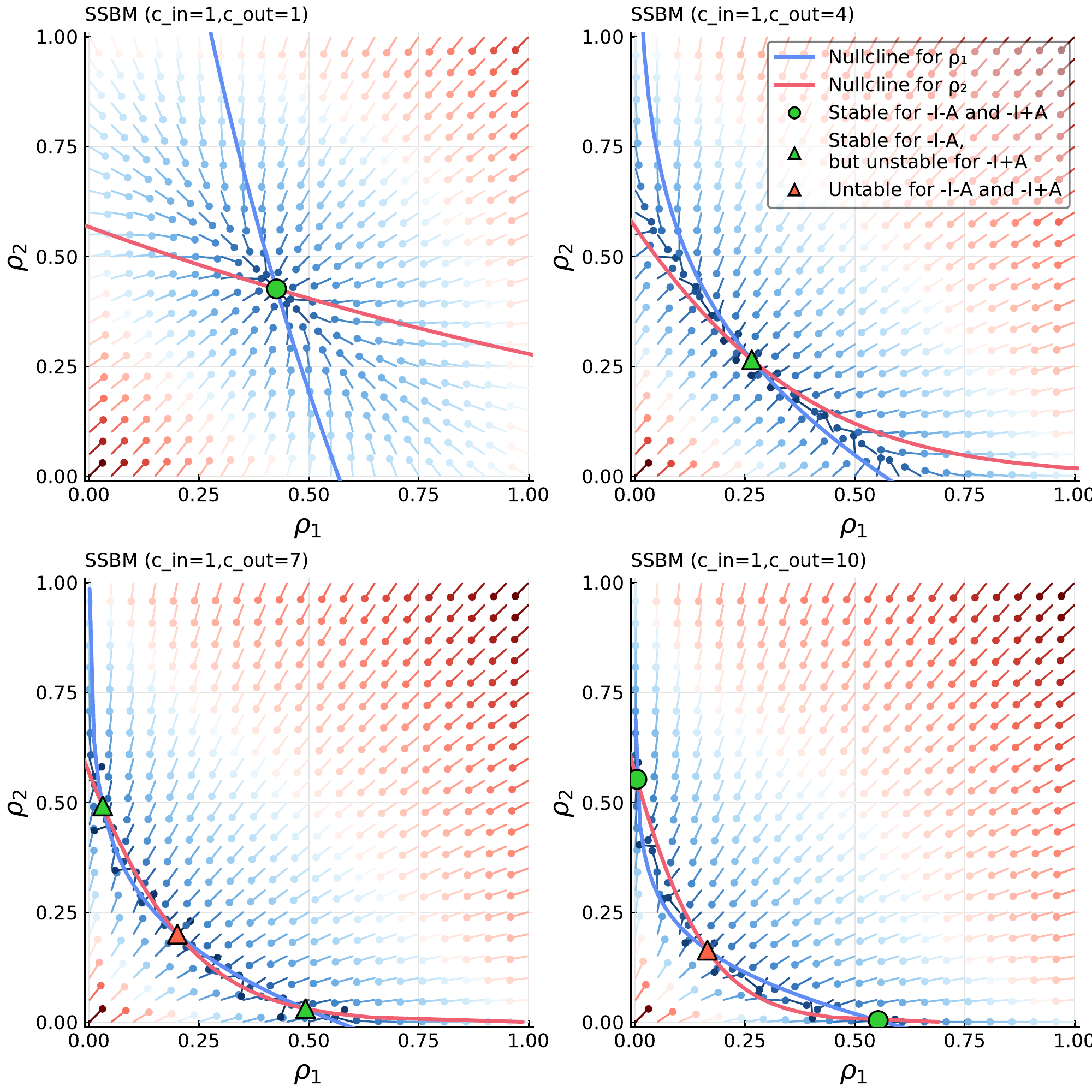}
    \caption{The phase planes of Eq.(\ref{eq:ssbm_rho_ev}) for SSBM with two communities. The mean degree of intra-community is fixed as $c_{\text{in}} = 1$. The description of this figure is the same as in Fig. \ref{fig:phaseplain1}.}
    \label{fig:phaseplain2}
\end{figure}

Figures \ref{fig:phaseplain1} and \ref{fig:phaseplain2} plot the phase planes of Eq.(\ref{eq:ssbm_rho_ev}) for SSBM with two communities, where fixed points are denoted by markers. The shape and color are determined by its stabilities, which are evaluated from the eigenvalues of $- I - A$ (Eq.(\ref{eq:25})) and $- I + A$ (Eq.(\ref{eq:26})). The former corresponds to the stability of the macroscopic variables $\rho=(\rho_1,\rho_2)^\top$ while the latter corresponds to the stability of the microscopic variables $\{u_{j\to i}\}$, respectively. 

In the case of $c_{\mathrm{in}} = 0$ (Fig. \ref{fig:phaseplain1}), at first there is only one stable fixed point. However, as $c_{\mathrm{out}}$ increases, a bifurcation occurs at $c_{\rm in }+c_{\rm out}=c_{\rm out}=e$, which yields two stable and one unstable fixed points. The stable points assign a value of one to most of the binary messages sent across the two communities. This indicates that either of the two communities is covered more when the bipartite structure becomes sufficiently strong by increasing $c_{\mathrm{out}}$.

The behavior for the case of $c_{\rm in}=1$ (Fig. \ref{fig:phaseplain2}) is similar to that of $c_{\rm in}=0$ for relatively small $c_{\rm out}$. Namely, a single stable fixed point bifurcates to two macroscopically stable and one macroscopically unstable fixed points when the total mean degree $c_{\rm in} + c_{\rm out} =1+ c_{\rm out}$ reaches $e$ from below. However, the resulting two macroscopically stable fixed points are unstable microscopically, which indicates that the messages continue to move while keeping the macroscopic distribution stationary. Interestingly, as $c_{\rm out}$ grows further, the macroscopically stable fixed points become microscopically stable as well at a certain critical point which is larger than $e$. This behavior can be interpreted to be a consequence of the emergence of strong bipartite structure created in SBM. As mentioned in Sec \ref{subsec:2-1}, Min-VCP belongs to the class P for bipartite graphs.

\begin{figure}[t]
    \centering
    \includegraphics[width=\hsize]{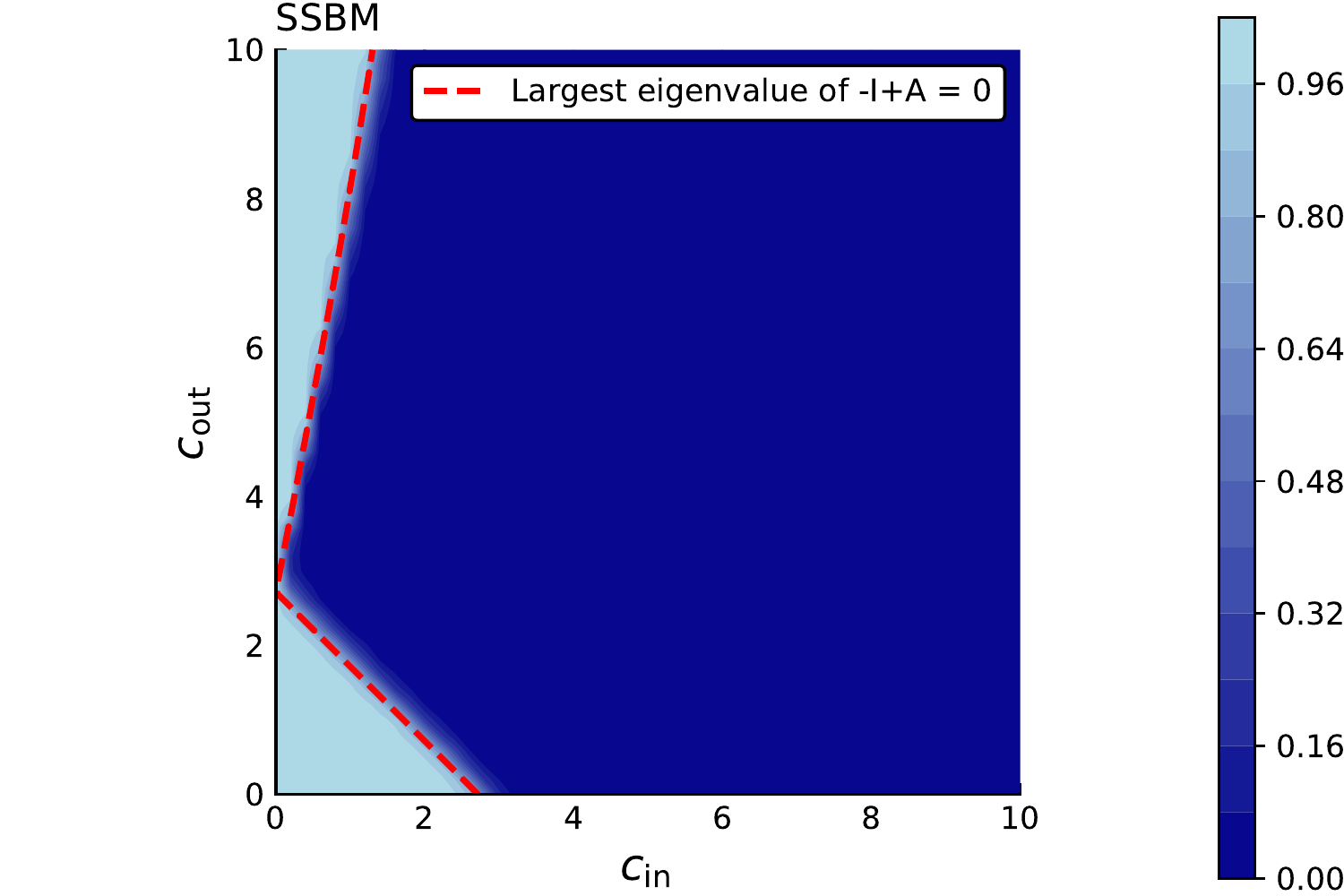}
    \caption{The convergence probability of WP for SSBM with two communities as a function of $c_{\mathrm{in}}$ and $c_{\mathrm{out}}$. The red dashed lines stand for the boundary at which the largest eigenvalue of $- I + A$ of Eq.(\ref{eq:26}) vanishes.}
    \label{fig:convergenceproba}
\end{figure}

To confirm these theoretical predictions, we assessed the probability that the sequential random update of WP converges. This probability is calculated according to a method of \cite{weigt2006message}. The result, in conjunction with the boundary at which the largest eigenvalue of $-I+A$ of Eq. (\ref{eq:26}) vanishes (red dashed line), is plotted in Fig. \ref{fig:convergenceproba}, which is quite consistent with the above mentioned scenario.

\begin{figure}[t]
    \centering
    \includegraphics[width=\hsize]{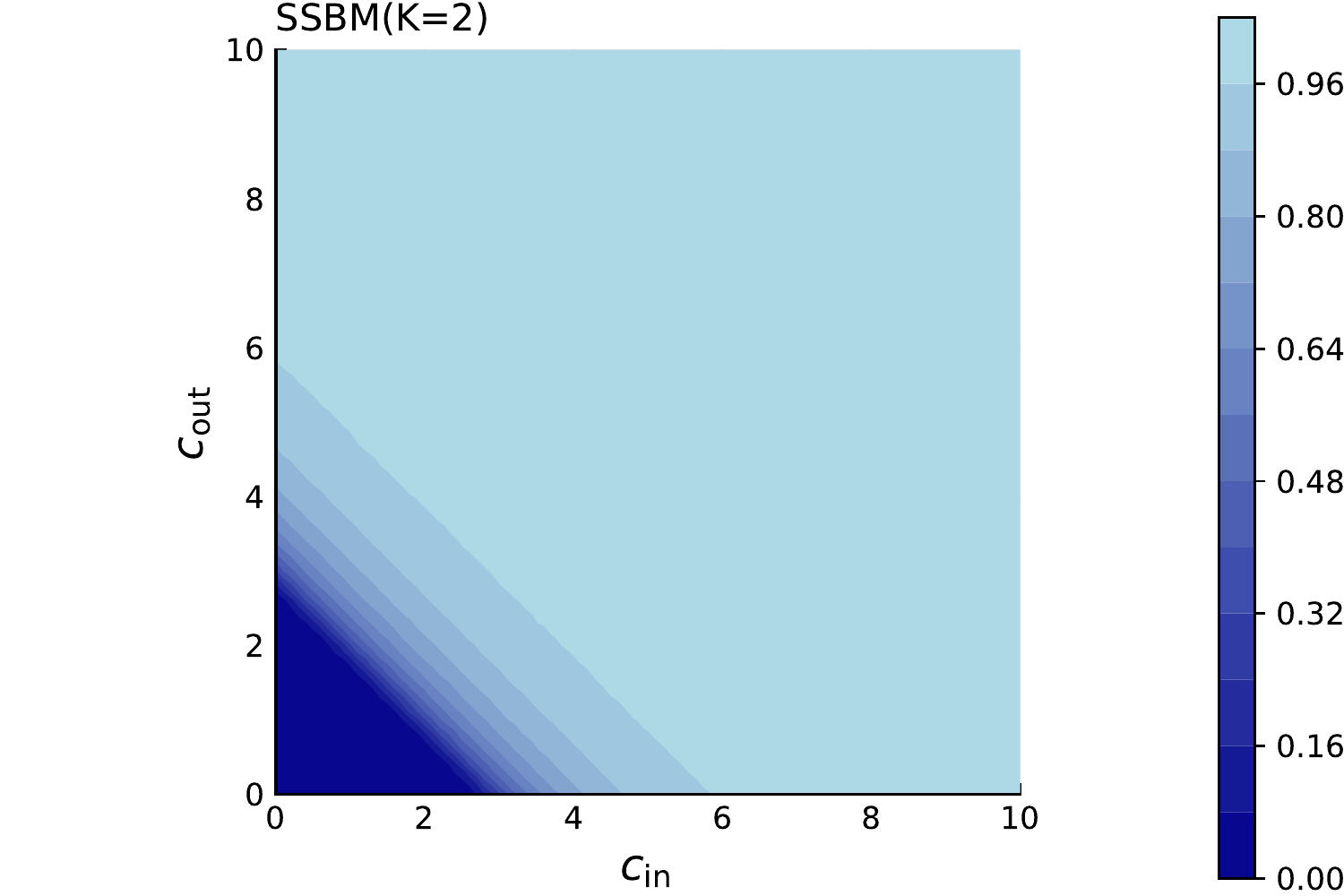}
    \caption{Core percolation: The ratio of the vertices that belong to the core for SSBM with two communities. The values are evaluated from 100 sample graphs with $N = 2500$.}
    \label{fig:corepercolation}
\end{figure}

In ERM, it is reported that the instability of WP is associated with the core percolation \cite{liu2012core}, \cite{azimi2018generalization}, \cite{zhao2018two}. To examine how this association is relevant in SBM, we plot the ratio of the vertices that belong to the core in Fig. \ref{fig:corepercolation}. This figure shows that although the critical condition of the core percolation is in accordance with that of the first transition at $c_{\rm in}+c_{\rm out}=e$, it is not relevant to the second (reentrant) transition that appears in the region of $c_{\rm out} > e$. This indicates that the core percolation is not the sole cause for the emergence of the computational difficulty, and the mesoscopic structures of graphs strongly influence it. 

\subsection{Experimental validation}

\begin{figure*}[htbp]
  \begin{center}
    \begin{tabular}{c}

      \begin{minipage}{0.5\hsize}
        \begin{center}
          \includegraphics[width=0.9\hsize]{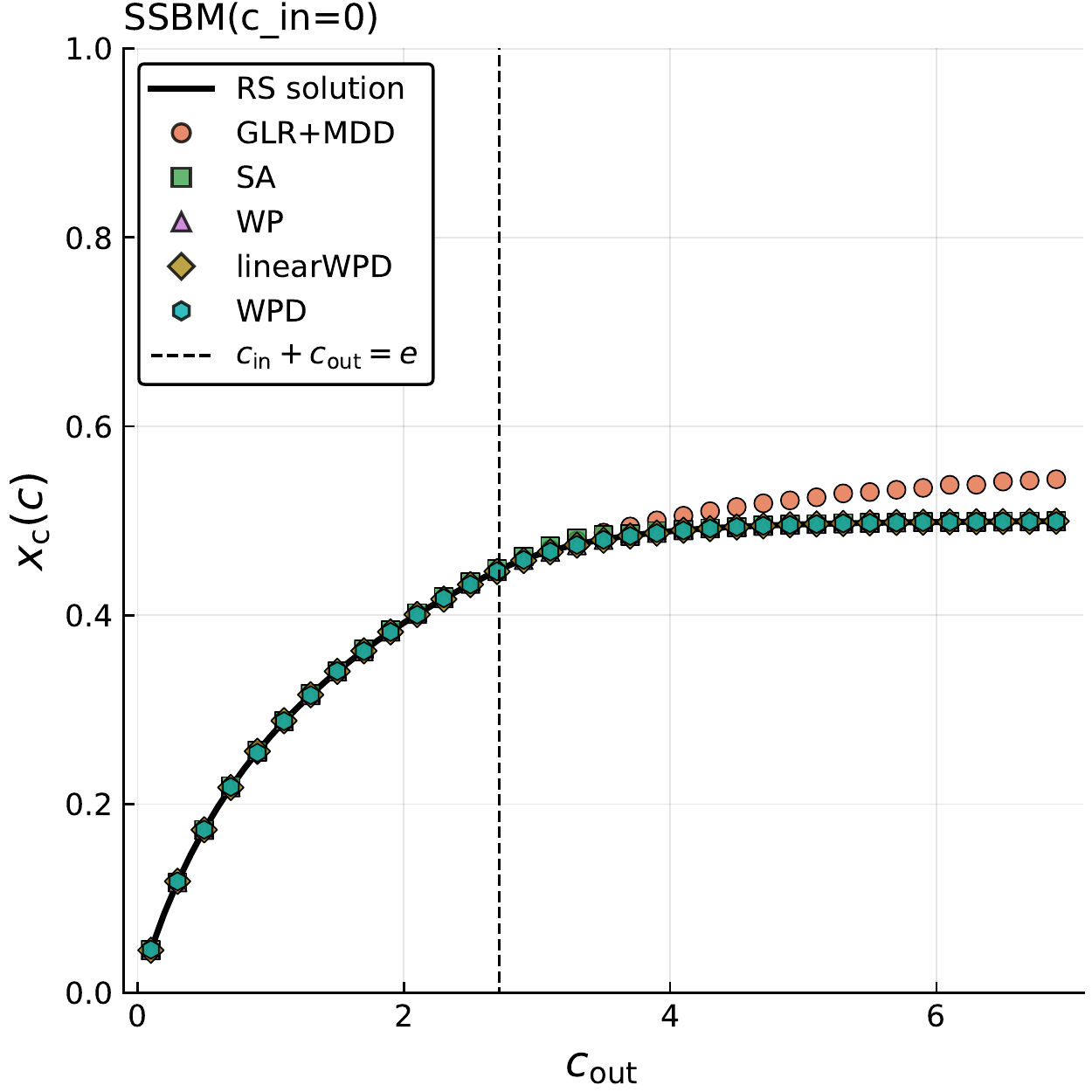}
        \end{center}
      \end{minipage}

      \begin{minipage}{0.5\hsize}
        \begin{center}
          \includegraphics[width=0.9\hsize]{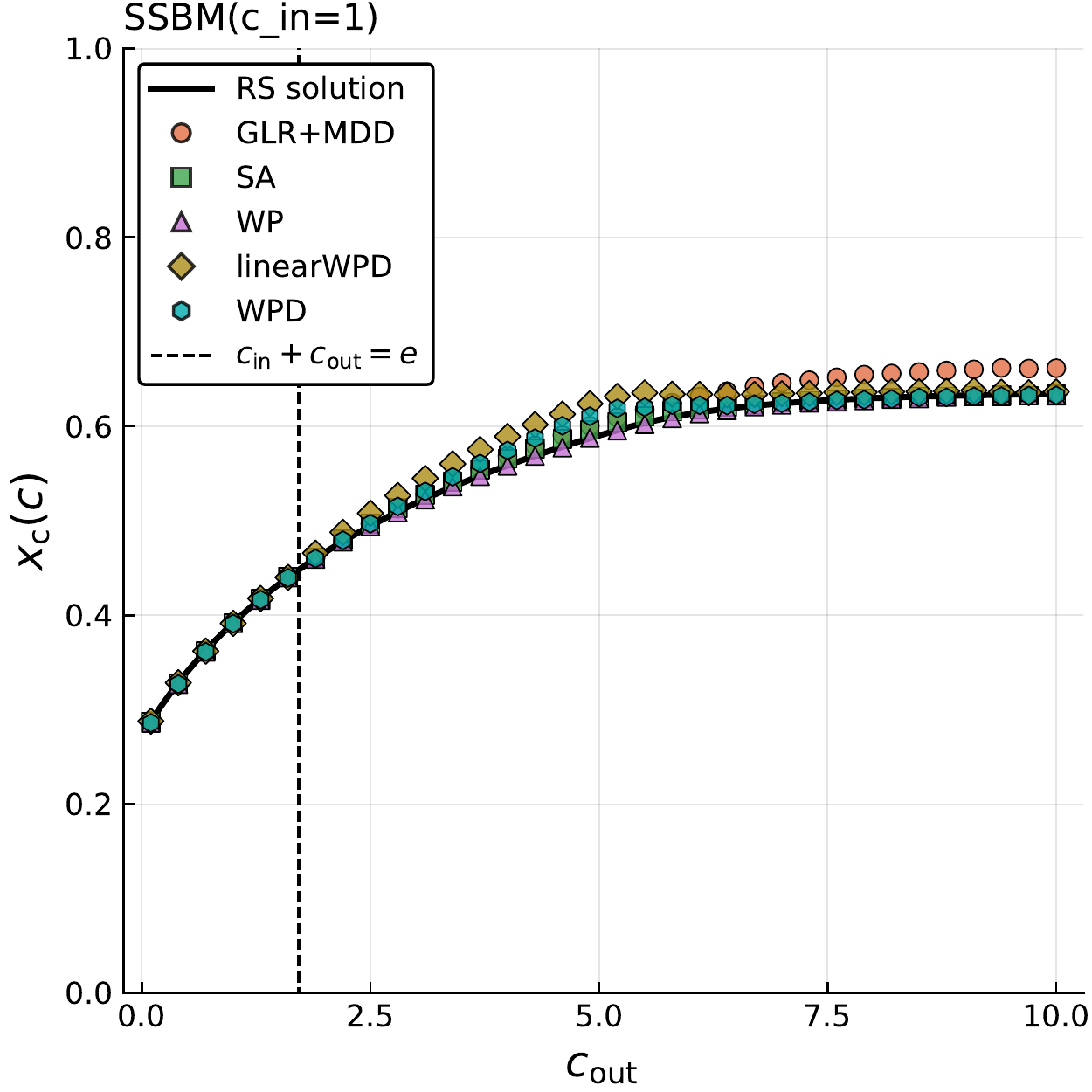}
        \end{center}
      \end{minipage}

    \end{tabular}
    \caption{The cover ratios of Min-VC for SSBM with two communities as a function of $c_{\mathrm{out}}$. Left and right panels correspond to the cases of $c_{\rm in}=0$ and $1$, respectively. Each point indicates the sample average obtained by the corresponding algorithm. Black solid curves represent the replica-symmetric solution obtained by Eq. (\ref{eq:de3}). Black dotted vertical lines are placed at the point of $c_{\mathrm{in}} + c_{\mathrm{out}} = e$.}
    \label{fig:coverratio}
  \end{center}
\end{figure*}

We performed WP on 100 instances of SSBM with $N = 5000$ vertices. In the experiments, we set the maximum number of updates for WP to $100 \times 2 M$ times, where $M$ is the number of edges. After WP converged or the number of updates reached the maximum, we calculated the cover ratio $x_{\mathrm c}$ using  Eq. (\ref{eq:15}). This is plotted in Fig. \ref{fig:coverratio} along with the analytical result (solid curve).

We also tested other algorithms to obtain $x_{\mathrm c}$ namely, the hybrid algorithm with greedy leaf removal and maximum degree decimation (GLR+MDD) \cite{jin2014statistical}, simulated annealing (SA) \cite{kirkpatrick1983optimization} \cite{lucas2014ising}, WP, WP decimation (WPD) \cite{jin2014statistical} and WP-based linear-time-and-space algorithm (linearWPD) \cite{xu2018warning}.

In the case of $c_{\mathrm{in}} = 0$, the results obtained from each algorithm are in agreement for $c_{\mathrm{out}} < e$. However, for $c_{\mathrm{out}} > e$, the result obtained by the GLR+MDD algorithm starts to deviate from the results of other algorithms. This can be understood as a consequence of the core-percolation which occurs at $c_{\mathrm{out}} = e$.

On the other hand, in the case of $c_{\mathrm{in}} = 1$, when $1 + c_{\mathrm{out}} > e$, the results from GLR+MDD algorithm as well as those from the WP-based algorithms start to deviate from the result of SA algorithm, which is regarded as the ground truth. However, when $c_{\mathrm{out}}$ becomes larger, the results of the WP-based algorithms agree with the result of SA again, supporting the existence of the reentrant transition predicted above.

\subsection{Visualization}

\begin{figure*}
    \centering
    \includegraphics[width=\hsize]{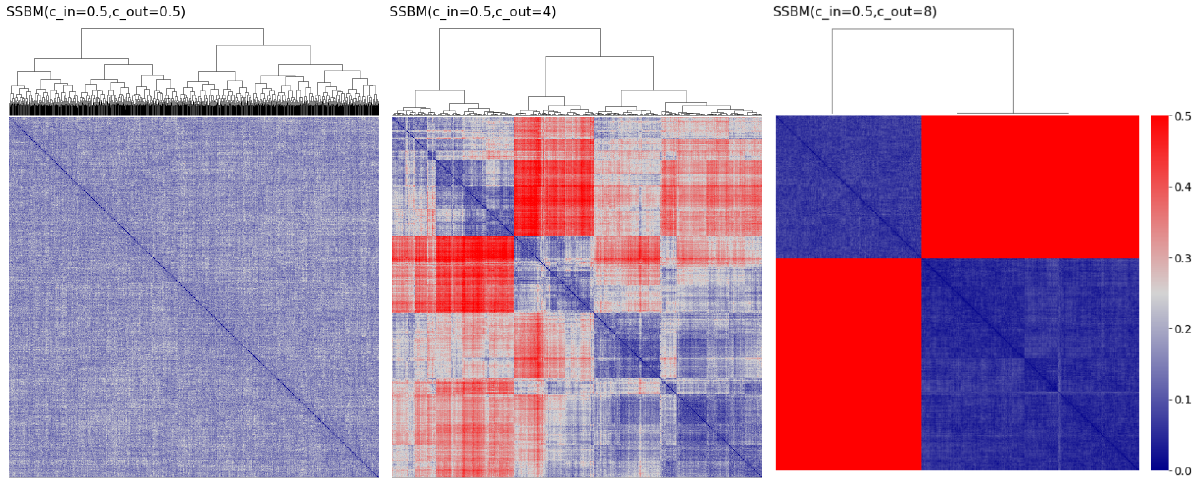}
    \caption{The heat map of Hamming distance matrix for 1024 Min-VC samples given by SA for SSBM with two communities of $N = 256$. Intra-degree $c_{\mathrm{in}}$ is fixed to 0.5. Left, center, and right panels correspond to the cases of $c_{\rm out}=0.5, 4,$ and $8$, respectively. The arrangement is determined by Ward's method and its hierarchical structure is indicated by the dendrogram.}
    \label{fig:dendogram}
\end{figure*}

Two visualization methods are used to explain the above results graphically. First, we use Ward's method \cite{ward1963hierarchical}, which is one of the hierarchical clustering methods. This method was applied to the solutions for Min-VCP in earlier studies \cite{hartmann2006phase}. We sampled 1024 solutions by SA and clustered them by Hamming distance. Fig. \ref{fig:dendogram} plots the dendrogram and heat map of the distance matrix. As $c_{\mathrm{out}}$ increases, a hierarchical structure comes out in the solution space, but the solutions are eventually aggregated into two points.

\begin{figure*}
    \centering
    \includegraphics[width=\hsize]{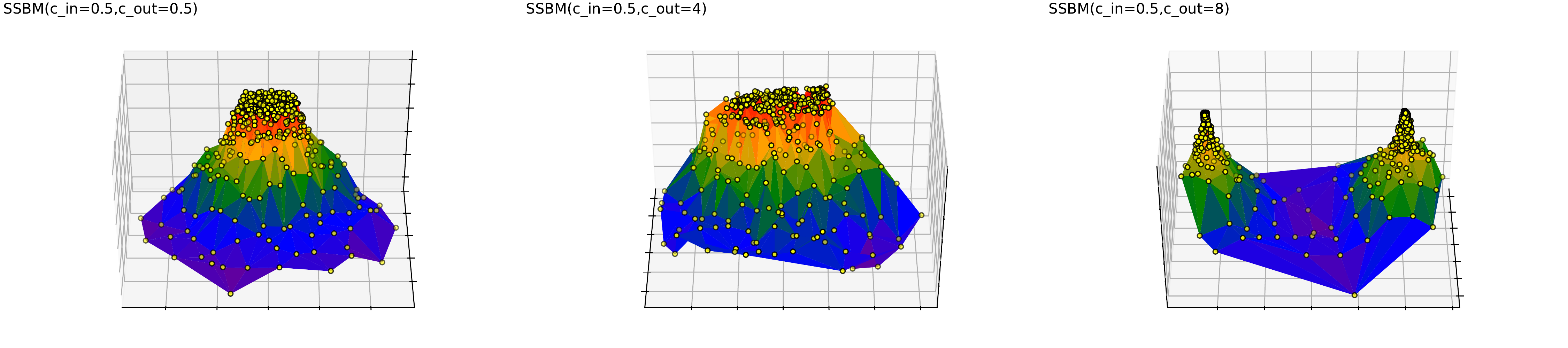}
    \caption{The landscape of 512 Min-VC samples given by SA for SSBM with two communities of $N = 256$. Intra-degree $c_{\mathrm{in}}$ is fixed to 0.5. Left, center, and right panels correspond to the cases of $c_{\rm out}=0.5, 4,$ and $8$, respectively. Each solution is mapped from $\{0, 1\}^N$ to $\mathbb R^2$ by curvilinear component analysis.}
    \label{fig:landscape}
\end{figure*}

We also use the curvilinear component analysis (CCA) \cite{demartines1997curvilinear}, which is one of the nonlinear dimensionality reduction algorithms. The papers \cite{good2010performance} \cite{kawamoto2019counting} applied CCA to the solutions of modularity maximization problem. In the current study, we applied CCA to the solutions of Min-VCP. Fig. \ref{fig:landscape} plots the landscape, where we can more visually confirm how the characteristic features of the solution space vary as $c_{\rm out}$ increases.

\section{\label{sec:5}Summary}

This study analyzed the typical property of the Min-VCP on SBM using the cavity method. In particular, we examined the critical condition of the computational difficulty for searching for Min-VC sets by the linear stability and bug proliferation analyses for WP. We also performed numerical experiments for SSBM with two communities by various algorithms, which supported predictions obtained by the theoretical analysis.

Min-VC sets are easily found by WP when the total mean degree $c_{\rm in}+c_{\rm out}$ is relatively small. However, it becomes difficult when $c_{\rm in}+c_{\rm out}$ exceeds $e=2.718...$. This transition is regarded as a consequence of the replica symmetry breaking caused by the core-percolation transition, which is also observed in ERM. However, when $c_{\mathrm{out}}$ becomes sufficiently larger than $c_{\mathrm{in}}$ in the region of $c_{\rm out}>e$, the solution search by WP becomes easy again. The Min-VCP on a bipartite graph is known to be equivalent to the maximum matching problem that belongs to class P by the K\"{o}nig's theorem. The reentrant behavior from the computationally difficult to easy phases presumably reflects this fact. We confirmed this by evaluating the convergence probability of WP and various numerical experiments. These indicate that mesoscopic structures such as ``communities'' strongly influence the computational difficulty for finding solutions of problems defined on graphs.

Future research scope includes extension of the current analysis to the hyper graphs, which corresponds to {\em hitting set problem} \cite{mezard2007statistical} \cite{takabe2014minimum} further investigation of RSB using survey propagation \cite{weigt2006message} \cite{zhang2009stability} and validation of the bug analysis \cite{castellani2005spin}.

\section*{Acknowledgements}

This work was partially supported by KAKENHI No. 17H00764 and a collaboration with Fujitsu Laboratories Ltd.

\nocite{*}

\end{document}